\documentstyle[IJFCS,QC,amsfonts,amsmath,amssymb]{article}

\long\def\magma#1{}

\def\ket#1{|#1\rangle}
\def\bra#1{\langle#1|}
\def\braket#1#2{\langle#1|#2\rangle}

\def\trace{\mathop{\rm tr}}
\def\wgt{\mathop{\rm wgt}}

\def\ADD{{\rm ADD}}
\def\DFT{{\rm DFT}}
\def\HORNER{{\rm HORNER}}

\def\C{{\mathbb{C}}}
\def\F{{\mathbb{F}}}
\def\Z{{\mathbb{Z}}}
\def\openone{id}

\let\ds\displaystyle

\makeatletter
\newcounter{lcnt}
\def\mylcnt{\hbox to 1em{\hfil\thelcnt}}
\def\thelcnt{\arabic{lcnt}}
\newenvironment{listing}{\setcounter{lcnt}{0}%
\def\cr{\\\stepcounter{lcnt}\mylcnt\>}\def\@currentlabel{\thelcnt}%
\begin{tabbing}\quad\quad\=\quad\=\quad\=\quad\=\quad\=\kill%
\stepcounter{lcnt}\mylcnt\>}{\end{tabbing}}%
\makeatother

\begin{document}
\copyrightheading

\symbolfootnote

\textlineskip

\begin{center}

\fcstitle{EFFICIENT QUANTUM CIRCUITS FOR NON-QUBIT\newline QUANTUM
ERROR-CORRECTING CODES} 

\vspace{24pt}

{\authorfont MARKUS GRASSL,\footnote{Current address: The Mathematical
        Sciences Research Institute, 
        1000 Centennial Drive, \#5070,
        Berkeley, CA 94720-5070, USA. E-mail: grassl@ira.uka.de}
MARTIN R\"OTTELER,\footnote{E-mail: roettele@ira.uka.de} and
THOMAS BETH\footnote{E-mail: EISS\_Office@ira.uka.de}}

\vspace{2pt}

\smalllineskip
{\addressfont Institut f\"ur Algorithmen und Kognitive Systeme,
        Universit\"at Karlsruhe,\\ Am Fasanengarten 5, 76\,128
        Karlsruhe, Germany.}
\vspace{20pt}

\def\today{November 4, 2002}
\publisher{(\today)}{}{Jozef Gruska}

\end{center}

\alphfootnote

\begin{abstract}
We present two methods for the construction of quantum circuits for
quantum error-correcting codes (QECC).  The underlying quantum systems
are tensor products of subsystems (qudits) of equal dimension which is
a prime power. For a QECC encoding $k$ qudits into $n$ qudits, the
resulting quantum circuit has $O(n(n-k))$ gates. The running time of
the classical algorithm to compute the quantum circuit is
$O(n(n-k)^2)$.

\end{abstract}

\textlineskip
\section{Introduction}
Most quantum error-correction codes (QECC) have been
constructed for quantum systems that are composed of two-dimensional
subsystems---quantum bits or short qubits. At first glance, the
size of the alphabet used to encode the information to be processed
seems to be irrelevant. But already in the context of classical
information processing and communication this is not true. For
example, the entropy of a message depends on the alphabet used for
the encoding, and increasing the size of the alphabet allows the
construction of better error-correcting codes \cite{MS77,TVZ82}.

In the context of quantum information, a quantum system consisting of
two three-dimensional subsystems---qutrits---shows new features
when compared to a two-qubit system. For example, bound entanglement
exists only for the former \cite{Hor97}, and for two qutrits there is
even ``nonlocality without entanglement'' \cite{BDF99}.

In this paper, we consider quantum systems which have subsystems of
dimension $d=p^m$, where $p$ is prime. As a shorthand, we will use the
term ``qudit''. Quantum codes for qudit systems have been studied,
e.\,g., in \cite{AhBO97,AsKn01,Got98:NASA,Rai99:nonbinary}.  The
question of encoding and decoding these codes, however, was not
explicitly addressed. Here we present efficient (classical) algorithms
to compute efficient quantum networks for the encoding process. The
method applies to qubit codes as well, for which encoding algorithms
have been discussed, e.\,g., in
\cite{ClGo97,Gra02,Gras02,GrBe00,GrGeBe99}.

In the following section, we present the mathematical framework of
quantum information processing using composed quantum systems. In
Section~\ref{sec:QECC} we recall the basic concepts and constructions
of quantum error-correcting codes. A first encoding algorithm for the
class of CSS codes is presented in Section~\ref{sec:CSS}. The main
result, an encoding algorithm for stabilizer codes over higher
dimensional quantum-systems, is derived in Section~\ref{sec:Jacobi} and
illustrated in Section~\ref{sec:Example}.

\section{Non-binary Quantum Systems}\label{sec:qudits}
\subsection{Quantum States and Registers}
In the context of classical information processing, information is
encoded using words over some finite alphabet ${\cal A}$, most often
${\cal A}=\{0,1\}$. The elementary ``symbols'' of quantum information
processing are states of a finite dimensional quantum system. Those
can be modeled by normalized vectors in a complex Hilbert space ${\cal
H}=\C^d$. Again, the simplest case of ${\cal H}=\C^2$ is considered
most often. Then the orthonormal basis states of such a quantum bit,
or short qubit, are written as $\ket{0}$ and $\ket{1}$. (The notation
of quantum states as ``ket vectors'' $\ket{\:\cdot\:}$ is attributed
to Dirac \cite{Dir58}).  A general state of a qubit is given by
$$
\ket{\phi}=\alpha\ket{0}+\beta\ket{1},\qquad
\text{where $\alpha,\beta\in\C$ and $|\alpha|^2+|\beta|^2=1$.}
$$
If both $\alpha$ and $\beta$ are non-zero, the state $\ket{\phi}$ is a
so-called superposition of $\ket{0}$ and $\ket{1}$ with amplitudes
$\alpha$ and $\beta$.  For a $d$-dimensional system, we label the
orthonormal basis states by the elements of some alphabet of size $d$,
e.\,g., the numbers $\{0,1,\ldots,d-1\}$ or the elements of a finite
field, if $d$ is a prime power. The general state of a qudit is given
by
$$
\ket{\psi}=\sum_{i=0}^{d-1}\alpha_i\ket{i},\qquad
\text{where $\alpha_i\in\C$ and $\sum_{i=0}^{d-1}|\alpha_i|^2=1$.}
$$
Combining several qudits, we obtain a quantum register. The canonical
basis states of a quantum register of length $n$ are tensor products
of the basis states of the single qudits. Hence we can label them by
words $x\in{\cal A}^n$ of length $n$. For the basis states of a
quantum register we use the following notations:
$$
\ket{x_1}\otimes\ket{x_2}\otimes\ldots\otimes\ket{x_n}
=\ket{x_1}\ket{x_2}\ldots\ket{x_n}=\ket{x_1,x_2,\ldots,x_n}=\ket{x}.
$$
A general state of a quantum register of length $n$ is a normalized
vector in the exponentially large Hilbert space ${\cal
H}=(\C^d)^{\otimes n}\cong \C^{d^n}$, given by
$$
\ket{\Psi}=\sum_{x\in{\cal A}^n}\alpha_x\ket{x},\qquad
\text{where $\alpha_x\in\C$ and $\sum_{x\in{\cal A}^n}|\alpha_x|^2=1$.}
$$
When writing states of quantum registers, normalization factors may be
omitted.

The elements of the dual space ${\cal H}^*$ will be denoted by ``bra
vectors'' $\bra{y}$. Then the inner product of two vectors $\ket{x}$
and $\ket{y}$ reads $\braket{x}{y}$, and linear operators $M$ can be
written in the form
$$
M=(M_{i,j})=\sum_{i,j}M_{ij}\ket{i}\bra{j}.
$$
As the operations have to preserve the normalization of the vectors,
the admissible operations are unitary operators, which we will discuss next.

\subsection{Elementary Gates}
In the following we will consider qudit systems where each qudit
corresponds to a $q$-dimensional Hilbert space where $q=p^m$ is a
prime power.
\begin{definition}[Elementary Gates]\label{def:elemGates}
Let $q$ be a prime power, i.\,e., $q=p^m$ where $p$ is prime. By
$\omega$ we denote a primitive complex $p$-th root of unity, i.\,e.,
$\omega=\exp(2\pi i/p)$. Furthermore, let $\trace(\alpha)$ denote the
trace of an element $\alpha\in\F_q=\F_{p^m}$ which is defined as
$\trace(\alpha):=\sum_{i=0}^{m-1}\alpha^{p^i}\in\F_p$. Then we define
the following operations:
\begin{romanlist}
\item $\ds X_\alpha:=\sum_{x\in\F_q}\ket{x+\alpha}\bra{x}$ for
$\alpha\in\F_q$\label{def:X}
\item $\ds Z_\beta:=\sum_{z\in\F_q}\omega^{\trace(\beta z)}\ket{z}\bra{z}$ for
$\beta\in\F_q$
\item $\ds M_\gamma:=\sum_{y\in\F_q}\ket{\gamma y}\bra{y}$ for $\gamma\in\F_q\setminus\{0\}$\label{def:mult}
\item $\ds\DFT:=\frac{1}{\sqrt{q}}\sum_{x,z\in\F_q}\omega^{\trace(xz)}\ket{z}\bra{x}$ \label{def:DFT}
\item $\ds\ADD^{(1,2)}:=\sum_{x,y\in\F_q}\ket{x}_1\ket{x+y}_2\bra{y}_2\bra{x}_1$\label{def:ADD}
\item $\ds\HORNER^{(1,2,3)}:=\sum_{a,x,b\in\F_q}
  \ket{a}_1\ket{x}_2\ket{ax+b}_3\bra{b}_3\bra{x}_2\bra{a}_1$
\end{romanlist}
\end{definition}
Here, when writing $\omega^{\trace(\beta z)}$, we identify $\F_p$ and
$\Z/p\Z$, the integers modulo $p$. Then the function $\chi_\beta\colon
z\mapsto \omega^{\trace(\beta z)}$ is an additive character of
$\F_q$. Different values of $\beta$ yield the $q$ different additive
characters of $\F_q$ (see, e.\,g., \cite{Jun93,LiNi83}).

\begin{figure}[hbt]
\unitlength1.2\unitlength
\centerline{
\OneQubitGate(1,3){$X_\alpha$}
\OneQubitGate(1,3){$Z_\beta$}
\OneQubitGate(1,3){$M_\gamma$}
\OneQubitGate(1,3){\scriptsize$\DFT$}
\CNOT(1,2,3)
\multCNOT[1,2](3,3)}
\centerline{
\makebox[30 \unitlength]{(i)}%
\makebox[30 \unitlength]{(ii)}%
\makebox[30 \unitlength]{(iii)}%
\makebox[30 \unitlength]{(iv)}%
\makebox[30 \unitlength]{(v)}%
\makebox[30 \unitlength]{(vi)}%
}
\fcaption{Graphical representation of the elementary gates of Definition~\ref{def:elemGates}.\label{fig:elem_gates}}
\end{figure}
A graphical representation of these elementary operations---so-called
quantum gates---is given in Figure~\ref{fig:elem_gates}. Each
horizontal line corresponds to one qudit. The first three operations
operate on single qudits. A superscript in brackets indicates on which
subsystem the transformation acts, e.\,g.,
$X_\alpha^{(2)}=\openone\otimes
X_\alpha\otimes\openone\otimes\ldots\otimes\openone$.  The operations
$X_\alpha$ and $M_\gamma$ correspond to the addition of a fixed
element $\alpha\in\F_q$ and the multiplication with a fixed element
$\gamma\ne0$, respectively. The operation $Z_\beta$ has no direct classical
analogue, it changes the phases of the basis states. The Fourier
transformation $\DFT$ can be used to transform the state $\ket{0}$ to
a superposition of all basis states with equal amplitudes, i.e.,
$$
\DFT\ket{0}=\frac{1}{\sqrt{q}}\sum_{\alpha\in\F_q}\ket{\alpha}.
$$ 
Starting from this superposition, a quantum computation can, e.\,g.,
evaluate a function in parallel for all possible inputs. Arbitrary
classical functions over $\F_q$ can be implemented using the gates
$\ADD^{(1,2)}$ and $\HORNER^{(1,2,3)}$. The former corresponds to the
reversible implementation of the addition of two elements. The first
qudit is called ``control'', the second ``target''. The latter is a
universal reversible gate over $\F_q$, as any function over $\F_q$
corresponds to a polynomial which can be evaluated using the Horner
scheme. It is the generalization of the so-called Toffoli gate
\cite{Toff80} for qubits, which is universal gate for reversible
boolean functions, as well as the Fredkin gate \cite{FrTo82}.

\section{Quantum Error-Correcting Codes}\label{sec:QECC}
\subsection{Unitary Error Bases}
In the following, we will recall the basic properties and
constructions of quantum error-correcting codes. In order to construct
an error-correcting code, one has to specify an error model. The error
model can be specified by a (finite) set ${\cal E}$ of error
operators. In \cite{KnLa97}, the following characterization of
error-correcting codes is given.
\begin{theorem}
Let ${\cal C}$ be a subspace of the Hilbert space ${\cal H}$ with
orthonormal basis $\{\ket{c_1},\ldots,\ket{c_K}\}$. Then ${\cal C}$ is
a quantum error-correcting code for the error-operators ${\cal
E}=\{E_1,\ldots, E_\mu\}$ if and only if there exists
$\alpha_{k,l}\in\C$ such that for all $\ket{c_i}$, $\ket{c_j}$ and
$E_k,E_l\in{\cal E}$
\begin{equation}\label{eq:KnillLaflamme}
\bra{c_i}E_k^\dagger E_l\ket{c_j}=\delta_{i,j}\alpha_{k,l}.
\end{equation}
\end{theorem}
Most quantum error-correcting codes are designed to correct local
errors, i.\,e., errors that effect only some subsystems. The error
acting on a single subsystem can be any linear transformation. It is
sufficient that condition (\ref{eq:KnillLaflamme}) holds for a basis
of the linear space of error operators. For qudit systems of prime
power dimension $q$, we consider the following set of unitary
operators:
\begin{equation}\label{eq:UEB}
{\cal E}=\{ X_\alpha Z_\beta\colon\alpha,\beta\in\F_q\}.
\end{equation}
It is not hard to show that those $q^2$ operators are an orthogonal
basis with respect to the inner product $\langle
A,B\rangle=\trace(A^\dagger B)$. Furthermore, they generate an error
group ${\cal G}_1$ of size $p q^2$ with center $\zeta({\cal
G}_1)=\langle\omega I\rangle$ (see
\cite{KlRo02:UEB1,Kni96_errorbase}). Any element of ${\cal G}_1$ can
uniquely be written as $\omega^\gamma X_\alpha Z_\beta$ where
$\gamma\in\{0,\ldots,p-1\}$ and $\alpha,\beta\in\F_q$. The commutation
relations of two elements are derived from
\begin{eqnarray*}
X_\alpha Z_\beta X_\alpha^{-1}
&=&\sum_{x\in\F_q}\ket{x+\alpha}\bra{x}
\sum_{z\in\F_q}\omega^{\trace(\beta z)}\ket{z}\bra{z}
\sum_{y\in\F_q}\ket{y}\bra{y+\alpha}\\
&=&\sum_{z\in\F_q}\omega^{\trace(\beta z)}\ket{z+\alpha}\bra{z+\alpha}\\
&=&\sum_{z\in\F_q}\omega^{\trace(\beta(z-\alpha))}\ket{z}\bra{z}\\
&=&\omega^{-\trace(\alpha\beta)}Z_\beta.
\end{eqnarray*}
Hence commuting two elements results in a phase factor, i.\,e.,
\begin{equation}\label{eq:commute}
(X_\alpha Z_\beta)(X_{\alpha'} Z_{\beta'})=
\omega^{\trace(\alpha'\beta-\alpha\beta')}(X_{\alpha'} Z_{\beta'})(X_\alpha Z_\beta).
\end{equation}
For an $n$-qudit system, the error basis and the error group are the
$n$-fold tensor products ${\cal E}^{\otimes n}$ and ${\cal G}_n:={\cal
G}_1^{\otimes n}$, respectively. The weight of an error-operator
$E\in{\cal E}^{\otimes n}$ is the number of tensor factors that are
different from identity. If a $K$-dimensional subspace ${\cal C}$ of
$(\C^q)^{\otimes n}$ can correct all errors of weight no greater than
$t$, ${\cal C}$ is a $t$-error-correcting code with minimum distance
$2t+1$, denoted by $((n,K,2t+1))$.

\subsection{Stabilizer Codes}
A particular class of quantum error-correcting codes are so-called
stabilizer codes (see \cite{AsKn01,CRSS98,Got96_hamming}). The basic
idea is to consider an Abelian subgroup ${\cal S}$ of the error group
${\cal G}_n$ such that its intersection with the center of ${\cal
G}_n$ is trivial. The stabilizer code ${\cal C}$ is defined as the
common eigenspace of the operators in ${\cal S}$. Stabilizer codes can
be described in terms of certain classical codes over finite fields.

Any element $E$ of the error group ${\cal G}_n$ can uniquely be
written as
$$
E=\omega^\gamma
(X_{\alpha_1}Z_{\beta_1})\otimes(X_{\alpha_2}Z_{\beta_2})\otimes\ldots\otimes(X_{\alpha_n}Z_{\beta_n})
=:\omega^\gamma X_\alpha Z_\beta,
$$
where $\gamma\in\{0,\ldots,p-1\}$ and
$\alpha=(\alpha_1,\alpha_2,\ldots,\alpha_n),
\beta=(\beta_1,\beta_2,\ldots,\beta_n)\in\F_q^n$. The weight of an
element $X_\alpha Z_\beta$ is the number of indices $i$ for which not
both $\alpha_i$ and $\beta_i$ are zero.  From the commutation relation
(\ref{eq:commute}), it follows that for $(\alpha,\beta),
(\alpha',\beta')\in \F_q^n\times\F_q^n$
$$
(X_\alpha Z_\beta)(X_{\alpha'}Z_{\beta'})=
\omega^{(\alpha,\beta)*(\alpha',\beta')}(X_{\alpha'}Z_{\beta'})(X_\alpha Z_\beta),
$$
where the inner product $*$ is defined by
\begin{equation}\label{eq:traceip}
(\alpha,\beta)*(\alpha',\beta'):=\sum_{i=1}^n 
\trace(\alpha_i'\beta_i-\alpha_i\beta_i').
\end{equation}
This shows that the group $\overline{{\cal G}}_n:={\cal G}_n/\langle
\omega I\rangle$ is isomorphic to
$\F_q^n\times\F_q^n$. Furthermore, two elements $X_\alpha Z_\beta$ and
$X_{\alpha'}Z_{\beta'}$ commute if and only if
$(\alpha,\beta)*(\alpha',\beta')=0$. Hence an Abelian subgroup ${\cal
S}$ of ${\cal G}_n$ corresponds to a subspace $C$ of
$\F_q^n\times\F_q^n$ that is contained in its dual $C^*$ with respect
to the inner product (\ref{eq:traceip}).
\begin{definition}[Stabilizer Matrix]\label{def:stab_mat}
Let ${\cal S}$ be an Abelian subgroup of ${\cal G}_n$ which has
trivial intersection with the center of ${\cal G}_n$. Furthermore, let
$\{g_1,g_2,\ldots,g_{n-k}\}$ where $g_i=\omega^{\gamma_i}
X_{\alpha_i}Z_{\beta_i}$ with $\gamma_i\in\{0,\ldots,p-1\}$ and
$(\alpha_i,\beta_i)\in\F_q^n\times\F_q^n$ be a minimal set of
generators for ${\cal S}$. Then a stabilizer matrix of the
corresponding stabilizer code ${\cal C}$ is a generator matrix of the
(classical) linear code $C\subseteq F_q^n\times F_q^n$. We will write
this matrix in the form
$$
\left(
\begin{array}{c|c}
\alpha_1 & \beta_1\\
\alpha_2 & \beta_2\\
\vdots & \vdots\\
\alpha_{n-k} & \beta_{n-k}
\end{array}
\right)\in\F_q^{(n-k)\times 2n}.
$$
\end{definition}
Any error operator $E$ that does not commute with
all elements $S\in{\cal S}$ will change the eigenvalue of an
eigenstate $\ket{\psi}$ of $S$ which can be detected by a
measurement. Recall that $E=X_\alpha Z_\beta$ does not commute with
all $S\in{\cal S}$ if and only if $(\alpha,\beta)\notin C^*$. Finally,
the minimum distance of a stabilizer code is the minimum weight of the
vectors $(\alpha,\beta)\in C^*\setminus C$, since the errors
corresponding to $C^*$ are those that cannot be detected, and the
operators corresponding to $C$ have no effect on the code.

\section{Encoding CSS Codes}\label{sec:CSS}
A special class of stabilizer codes are so-called CSS codes named
after Calderbank, Shor \cite{CaSh96} and Steane
\cite{Ste96:error}. Originally, they have been designed for qubit
systems, but they can easily generalized to any dimension (see,
e.\,g., \cite{AhBO97}). Given two linear codes $C_1=[n,k_1,d_1]_q$ and
$C_2=[n,k_2,d_2]_q$ over $\F_q$ with $C_2^\bot\subseteq C_1$, the
basis states of the corresponding CSS code are
\begin{equation}\label{eq:CSSstate}
\ket{\psi_{w}}:=\frac{1}{\sqrt{|C_2^\bot|}}\sum_{c\in
C_2^\bot}\ket{c+w},\qquad\text{where $w\in C_1$.}
\end{equation}
Two states $\ket{\psi_{w}}$ and $\ket{\psi_{w'}}$ are identical if and
only if $w-w'\in C_2^\bot$, otherwise they are orthogonal. As the
dimension of $C_2^\bot$ is $n-k_2$, there are $q^k$ different cosets
where $k=k_1+k_2-n$. Hence the dimension of the CSS code is $q^k$. Its
minimum distance is $d\ge\min(d_1,d_2)$. 

We illustrate the encoding of a CSS code for the code $[\![7,3,3]\!]_8$
over seven $8$-dimensional quantum systems. Let $\alpha\in\F_8$ be a
primitive element of $\F_8$ with minimal polynomial
$\mu_\alpha(X)=X^3+X+1$. The code $C=[7,2,6]_8$ with generator matrix
$$
G=
\begin{pmatrix}
1&   0& \alpha^3&   1& \alpha^3& \alpha  & \alpha\\
0&   1& \alpha^4&   1& \alpha^5& \alpha^5& \alpha^4
\end{pmatrix}
$$
is contained in its dual $C^\bot=[7,5,3]$ with generator matrix
$$
H=
\left(
\begin{array}{*7{c}}
1&   0& \alpha^3&   1& \alpha^3& \alpha  & \alpha\\
0&   1& \alpha^4&   1& \alpha^5& \alpha^5& \alpha^4\\
\hline
0&   0&	  1&   0&   0& \alpha^3& \alpha^5\\
0&   0&	  0&   1&   0&	 \alpha& \alpha^5\\
0&   0&	  0&   0&   1&	 \alpha& \alpha^4
\end{array}
\right).
$$
The first two rows of $H$ are equal to $G$. The cosets of $C^\bot/C$
are given by linear combinations of the last three rows. Denoting the
rows of $H$ by $h_1,h_2,\ldots,h_5$, we can rewrite equation
(\ref{eq:CSSstate}) as
$$
\ket{\psi_{a,b,c}}=
\frac{1}{\sqrt{|C|}}\sum_{i,j\in\F_8}
\ket{i h_1+j h_2+a h_3+b h_4+c h_5},
$$
where $a,b,c\in\F_8$. Applying a Fourier transformation to the first
and second qudit of the initial state
$\ket{00}\ket{a}\ket{b}\ket{c}\ket{00}$, we obtain
\begin{equation}\label{eq:example_DFT}
\frac{1}{8}\sum_{i,j\in\F_8}\ket{i}\ket{j}\ket{a}\ket{b}\ket{c}\ket{00}.
\end{equation}
Now we sequentially add the corresponding multiple of the rows of $H$
in reverse order, i.\,e., starting with $h_5$. As $H$ is in row
echelon form, this corresponds to a sequence of modified $\ADD$
gates---instead of adding the first qudit to the second, we have to
add a multiple of it. The resulting encoding circuit is shown in
Figure~\ref{fig:CSSencoding}.
\begin{figure}[hbt]
\centerline{\tiny\unitlength0.7\unitlength
\def\DFT{F}
\gdef\cnot(#1,#2,#3){\kern-5\unitlength\CNOT(#1,#2,#3)\kern-5\unitlength}
\inputwires[$\ket{0}$,$\ket{0}$,,$\ket{\phi_{\text{in}}}\left\{\rule{0pt}{30\unitlength}\right.$\kern-2mm,,$\ket{0}$,$\ket{0}$](7)
\rlap{\OneQubitGate[6](1,1){$\DFT$}}%
\rlap{\OneQubitGate[5](1,1){$\DFT$}}%
\rlap{\OneQubitGate[1](4,4){$\alpha^{\text{-1}}$}}%
\OneQubitGate(1,1){$\alpha^{\text{-4}}$}
\rlap{\cnot(5,6,7)}%
\cnot(5,7,7)%
\rlap{\OneQubitGate[1](6,6){$\alpha$}}%
\OneQubitGate(1,1){${\alpha^4}$}
\rlap{\OneQubitGate[1](6,6){$\alpha^{\text{-1}}$}}%
\OneQubitGate(1,1){$\alpha^{\text{-5}}$}
\rlap{\cnot(4,6,7)}%
\cnot(4,7,7)%
\rlap{\OneQubitGate[1](6,6){$\alpha$}}%
\OneQubitGate(1,1){$\alpha^5$}
\rlap{\OneQubitGate[1](6,6){$\alpha^{\text{-3}}$}}%
\OneQubitGate(1,1){$\alpha^{\text{-5}}$}
\rlap{\cnot(3,6,7)}%
\cnot(3,7,7)%
\rlap{\OneQubitGate[1](6,6){$\alpha^3$}}%
\OneQubitGate(1,1){$\alpha^5$}
\rlap{\OneQubitGate[4](3,3){$\alpha^{\text{-4}}$}}%
\rlap{\OneQubitGate[2](2,2){$\alpha^{\text{-5}}$}}%
\rlap{\OneQubitGate[1](1,1){$\alpha^{\text{-5}}$}}%
\OneQubitGate(1,1){$\alpha^{\text{-4}}$}
\rlap{\cnot(2,3,7)}%
\rlap{\cnot(2,4,7)}%
\rlap{\cnot(2,5,7)}%
\rlap{\cnot(2,6,7)}%
\cnot(2,7,7)%
\rlap{\OneQubitGate[4](3,3){$\alpha^4$}}%
\rlap{\OneQubitGate[2](2,2){$\alpha^5$}}%
\rlap{\OneQubitGate[1](1,1){$\alpha^5$}}%
\OneQubitGate(1,1){$\alpha^4$}
\rlap{\OneQubitGate[4](3,3){$\alpha^{\text{-3}}$}}%
\rlap{\OneQubitGate[2](2,2){$\alpha^{\text{-3}}$}}%
\rlap{\OneQubitGate[1](1,1){$\alpha^{\text{-1}}$}}%
\OneQubitGate(1,1){$\alpha^{\text{-1}}$}
\rlap{\cnot(1,3,7)}%
\rlap{\cnot(1,4,7)}%
\rlap{\cnot(1,5,7)}%
\rlap{\cnot(1,6,7)}%
\cnot(1,7,7)%
\rlap{\OneQubitGate[4](3,3){$\alpha^3$}}%
\rlap{\OneQubitGate[2](2,2){$\alpha^3$}}%
\rlap{\OneQubitGate[1](1,1){$\alpha$}}%
\OneQubitGate(1,1){$\alpha$}
\outputwires[,,,$\kern-2mm\left.\rule{0pt}{70\unitlength}\right\}\ket{\phi_{\text{enc}}}$](7) 
\qquad
} 
\medskip
\fcaption{Encoding circuit for the CSS code
$[\![7,3,3]\!]_8$. The Fourier transformation $\DFT$ is abbreviated by
$F$, and for the gate $M_\alpha$ we give only the value of
$\alpha$.\label{fig:CSSencoding}}
\end{figure}

For each row $h_k$ of $H$, we have a sequence of gates $\ADD^{(k,l)}$
corresponding to the non-zero entries $H_{kl}$ for $k\ne l$. The gate
$\ADD^{(k,l)}$ is conjugated by a multiplication gate $M_\gamma$ where
$\gamma=H_{kl}$. Note that we have not combined adjacent
multiplication gates. In general, $G$ is a generator matrix of
$C_2^\bot\subseteq C_1$, and $H$ is its completion to a generator
matrix of $C_1$. Using the construction illustrated above, we get
\begin{proposition}
Let $C_1=[n,k_1,d_1]_q$ and $C_2=[n,k_2,d_2]_q$ be linear codes over
$\F_q$ with $C_2^\bot\subseteq C_1$. Then there exists a quantum
circuit to encode the resulting CSS code ${\cal C}=[\![n,k,d]\!]_q$
over qudits of dimension $q$, where $k=k_1+k_2-n$, using $n-k_2$
$\DFT$ gates, at most $A:=k_1 n-\tbinom{k_1+1}{2}$ $\ADD$ gates, and at
most $A+(n-1)$ multiplication gates.
\end{proposition}
\proof{%
To create a state similar to (\ref{eq:example_DFT}), we need $\dim
C_2^\bot=n-k_2$ $\DFT$ gates. A generator matrix $H$ for $C_1$ has
$k_1$ rows. Hence the number $A$ of $\ADD$ gates is at most $k_1
n$. As $H$ can be chosen to be in row echelon form, we have $k_1$
entries that are one. Those correspond to the control
qudits. Furthermore, at least $\binom{k_1}{2}$ entries of $H$ are
zero, reducing the number of $\ADD$ gates further. Finally, each
$\ADD$ gate is conjugated by a multiplication gate. Adjacent
multiplication gates between the $\ADD$ gates can be combined, so we
count only the multiplication gates before the $\ADD$ gates, plus at
most $n-1$ multiplication gates at the end.
}
As the r{\^o}le of $C_1$ and $C_2$ in the construction of CSS codes is
completely symmetric, one has the freedom to choose the matrices $G$
and $H$ to be generator matrices of either $C_2^\bot$ and $C_1$,
respectively $C_1^\bot$ and $C_2$. This may reduce the number of gates
by a constant factor, as well as optimizing the particular generator
matrices.

\section{Encoding Qudit Stabilizer Codes}\label{sec:Jacobi}
In this section we derive an encoding algorithm for general stabilizer
codes over qudit systems of prime power dimension $q=p^m$. The main
idea is to transform the Abelian stabilizer group ${\cal
S}\subseteq{\cal G}_n$ of the stabilizer code ${\cal
C}=[\![n,k,d]\!]_q$ into a stabilizer group ${\cal S}_0$ for which
encoding is particularly easy. Up to a permutation $\pi\in S_n$ of the
qudits---which we may ignore in the sequel---, that group is ${\cal
S}_0:=\langle Z_1^{(1)},Z_1^{(2)},\ldots,Z_1^{(n-k)}\rangle$.  The
corresponding stabilizer code ${\cal C}_0$ cannot correct single qudit
errors at arbitrary positions as the common eigenstates of ${\cal
S}_0$ are tensor products of the fixed $(n-k)$-qudit state
$\ket{00\ldots0}$ and the unencoded state
$\ket{\phi_{\text{in}}}\in(\C^d)^{\otimes k}$, i.\,e., they are of the
form
\begin{equation}\label{eq:Code0}
\ket{00\ldots 0}\ket{\phi_{\text{in}}}\in(\C^d)^{\otimes n}.
\end{equation}
As the stabilizer groups ${\cal S}$ and ${\cal S}_0$ are conjugated to
each other there exists a transformation $D$ such that
\begin{equation}\label{eq:conjugation}
D^{-1} {\cal S} D=_{\pi} 
\langle Z^{(1)},Z^{(2)},\ldots,Z^{(n-k)}\rangle={\cal S}_0,
\end{equation}
where $=_{\pi}$ denotes equality up to the permutation $\pi$ of the
qudits. Combining (\ref{eq:Code0}) and (\ref{eq:conjugation}), it
follows that the states
$D^{-1}(\ket{00\ldots0}\ket{\phi_{\text{in}}})$ are common eigenstates
of ${\cal S}$. Hence the transformation $D^{-1}$ can be used to encode
that states of ${\cal C}$. In the following, we show that the
transformation $D$ can be efficiently decomposed into single qudit
transformations and $\ADD$-gates.

\subsection{Conjugating the Error Basis}
First we consider the operation of the Fourier transformation over the
additive group of the field $\F_q$ (see
Definition~\ref{def:elemGates}~\ref{def:DFT}) on the error group
modulo the center. The action on $Z_\beta$ is given by
\begin{eqnarray*}
\DFT^{-1} Z_\beta\DFT&=&
\frac{1}{\sqrt{q}}\sum_{i,j\in\F_q}\omega^{-\trace(ij)}\ket{i}\bra{j}
\sum_{z\in\F_q}\omega^{\trace(\beta z)}\ket{z}\bra{z}
\frac{1}{\sqrt{q}}\sum_{k,l\in\F_q}\omega^{\trace(kl)}\ket{k}\bra{l}\\
&=&
\frac{1}{q}\sum_{i,l\in\F_q}\sum_{j\in\F_q}^{d-1}
\omega^{-\trace(ij)}\omega^{\trace(\beta
j)}\omega^{\trace(jl)}\ket{i}\bra{l}\\
&=&
\frac{1}{q}\sum_{i,l\in\F_q}\sum_{j\in\F_q}
\omega^{\trace((l-i+\beta)j)}\ket{i}\bra{l}\\
&=&\sum_{l\in\F_q}\ket{l+\beta}\bra{l}=X_\beta.
\end{eqnarray*}
Similarly,
\begin{eqnarray*}
\DFT^{-1} X_\alpha\DFT&=&
\frac{1}{\sqrt{q}}\sum_{i,j\in\F_q}\omega^{-\trace(ij)}\ket{i}\bra{j}
\sum_{x\in\F_q}\ket{x+\alpha}\bra{x}
\frac{1}{\sqrt{q}}\sum_{k,l\in\F_q}\omega^{\trace(kl)}\ket{k}\bra{l}\\
&=&
\frac{1}{q}\sum_{i,l\in\F_q}
\sum_{x\in\F_q}
\omega^{-\trace(i(x+\alpha))}\omega^{\trace(xl)}\ket{i}\bra{l}\\
&=&
\frac{1}{q}\sum_{i,l\in\F_q}
\omega^{\trace(-i\alpha)}\sum_{x\in\F_q}
\omega^{\trace(x(l-i))}\ket{i}\bra{l}\\
&=&
\sum_{i\in\F_q}
\omega^{\trace(-i\alpha)}\ket{i}\bra{i}=Z_{-\alpha}
\end{eqnarray*}
As any element of $\overline{{\cal G}}_1$ corresponds to a row vector
$(\alpha,\beta)$, we can describe the action of $\DFT$ on
$\overline{{\cal G}}_1$ as the linear transformation
$\overline{\DFT}:=\left(\begin{smallmatrix}0&-1\\1&0\end{smallmatrix}\right)$.

Next we investigate the action of the matrix $M_\gamma$ corresponding
to the multiplication with $\gamma\in\F_q$, $\gamma\ne 0$ (see
Definition~\ref{def:elemGates}~\ref{def:DFT}). We compute
\begin{eqnarray*}
M_\gamma^{-1}X_\alpha Z_\beta M_\gamma
&=&\sum_{y\in\F_q}\ket{y}\bra{\gamma y}
\sum_{x\in\F_q}\ket{x+\alpha}\bra{x}
\sum_{z\in\F_q}\omega^{\trace(\beta z)}\ket{z}\bra{z}
\sum_{v\in\F_q}\ket{\gamma v}\bra{v}\\
&=&\sum_{y\in\F_q}\ket{y}\bra{\gamma y}
\sum_{x\in\F_q}\ket{x+\alpha}\bra{x}
\sum_{v\in\F_q}\omega^{\trace(\beta \gamma v)}\ket{\gamma v}\bra{v}\\
&=&\sum_{y\in\F_q}\ket{\gamma^{-1}y}\bra{ y}
\sum_{v\in\F_q}\omega^{\trace(\beta \gamma v)}\ket{\gamma
v+\alpha}\bra{v}\\ &=& \sum_{v\in\F_q}\omega^{\trace(\beta \gamma
v)}\ket{v+\gamma^{-1}\alpha}\bra{v}\\ &=&
\sum_{x\in\F_q}\ket{x+\gamma^{-1}\alpha}\bra{x}
\sum_{z\in\F_q}\omega^{\trace(\beta \gamma z)}\ket{z}\bra{z}\\
&=&X_{\gamma^{-1}\alpha} Z_{\gamma\beta}.
\end{eqnarray*}
Hence, $M_\gamma$ acts on $(\alpha,\beta)$ as
$\overline{M}_\gamma:=\left(\begin{smallmatrix}\gamma^{-1}&0\\0&\gamma\end{smallmatrix}\right)$.  

For the next operation, we have to distinguish the cases whether $q$
is odd or $q$ is even. If $q$ is odd, we define the operator
\begin{equation}\label{def:P_odd}
P_\gamma:=\sum_{y\in\F_q}\omega^{-\trace(\frac{1}{2}\gamma
y^2)}\ket{y}\bra{y}
\end{equation}
which commutes with $Z_\beta$. Furthermore, $X_\alpha$ acts on
$P_\gamma^{-1}$ as follows:
\begin{eqnarray*}
X_\alpha^{-1} P_\gamma^{-1} X_\alpha
&=&
\sum_{x\in\F_q}\ket{x}\bra{x+\alpha}
\sum_{y\in\F_q}\omega^{\trace(\frac{1}{2}\gamma y^2)}\ket{y}\bra{y}
\sum_{z\in\F_q}\ket{z+\alpha}\bra{z}\\
&=&
\sum_{y\in\F_q}\omega^{\trace(\frac{1}{2}\gamma (y+\alpha)^2)}\ket{y}\bra{y}\\
&=&
\omega^{\trace(\frac{1}{2}\gamma \alpha^2)}
\sum_{z\in\F_q}\omega^{\trace(\gamma \alpha z)}\ket{z}\bra{z}
\sum_{y\in\F_q}\omega^{\trace(\frac{1}{2}\gamma y^2)}\ket{y}\bra{y}\\
&=&
\omega^{\trace(\frac{1}{2}\gamma \alpha^2)}
 Z_{\gamma\alpha} P_\gamma^{-1}.
\end{eqnarray*}
Equivalently, we get
$$
P_\gamma^{-1} X_\alpha P_\gamma = \omega^{\trace(\frac{1}{2}\gamma
\alpha^2)} X_\alpha Z_{\gamma\alpha}.
$$
Hence $P_\gamma$ acts on $(\alpha,\beta)$ as
$\overline{P}_\gamma:=\left(\begin{smallmatrix}1&\gamma\\0&1\end{smallmatrix}\right)$. 

If $q$ is even, we have to use another definition for the operator
$P_\gamma$ as we cannot divide by two. For this, we fix an arbitrary
self-dual basis $B=\{b_1,\ldots,b_m\}$ of $\F_q=\F_{2^m}$ over $\F_2$
(see e.\,g., \cite{Jun93}). Hence, by definition, $\trace(b_i
b_j)=\delta_{i,j}$. Any element $\alpha\in\F_q$ can uniquely be
written as $\sum_{i=1}^m \alpha_i b_i$ where $\alpha_i\in\F_2$. The
coefficients $\alpha_i$ are given by $\alpha_i=\trace(b_i\alpha)$
since
$$
\trace(b_i\alpha)=\trace(b_i \sum_{j=1}^m \alpha_j b_j)
=\sum_{j=1}^m \alpha_j \trace(b_i b_j)
=\sum_{j=1}^m \alpha_j \delta_{i,j}=\alpha_i.
$$
\begin{lemma}
Let $q=2^m$ and let $B=\{b_1,\ldots,b_m\}$ be an arbitrary self-dual
basis of $\F_q$ over $\F_2$. Furthermore, we define an integer-valued
function on $\F_q$ as
$$
\wgt\colon\F_q\rightarrow 
    \Z, \alpha\mapsto |\{j:j \in \{1,2,\ldots, m\}|\trace(\alpha b_j)\ne 0\}|.
$$
Then the following holds for all $\alpha,y\in\F_q$:
$$
i^{\wgt (y+\alpha)} = i^{\wgt(\alpha)} i^{\wgt(y)} (-1)^{\trace(\alpha
y)},
$$
where $i\in\C$ with $i^2=-1$.
\end{lemma}
\proof{%
First we observe that $\wgt(y+\alpha)=\wgt(y)+\wgt(\alpha) -
2|\{j: y_j=\alpha_j=1\}|$. Second, the size of this set modulo two is
given by
$$
\trace(\alpha y)=\trace\left(
\biggl(\sum_{j=1}^m \alpha_j b_j\biggr)
\biggl(\sum_{k=1}^m y_k b_k\biggr)\right)
=\sum_{j,k=1}^m \alpha_j y_k \trace(b_j b_k)
=\sum_{j=1}^m \alpha_j y_j,
$$
which completes the proof.
}

Now we are ready to define an operator $P_1$ as
$$
P_1:=\sum_{y\in\F_q} (-i)^{\wgt(y)} \ket{y}\bra{y}=
\sum_{y\in\F_q} \prod_{j=1}^m (-i)^{\trace(y b_i)} \ket{y}\bra{y}.
$$
Again, $P_1$ commutes with all matrices $Z_\beta$. The action on
$X_\alpha$ is derived from
\begin{eqnarray*}
X_\alpha^{-1} P_1^{-1} X_\alpha
&=&
\sum_{x\in\F_q}\ket{x}\bra{x+\alpha}
\sum_{y\in\F_q}i^{\wgt(y)}\ket{y}\bra{y}
\sum_{z\in\F_q}\ket{z+\alpha}\bra{z}\\
&=&
\sum_{y\in\F_q}i^{\wgt(y+\alpha)}\ket{y}\bra{y}\\
&=&
i^{\wgt(\alpha)}
\sum_{z\in\F_q}(-1)^{\trace(\alpha z)}\ket{z}\bra{z}
\sum_{y\in\F_q}i^{\wgt(y)}\ket{y}\bra{y}\\
&=&
i^{\wgt(\alpha)}
Z_{\alpha} P_1^{-1}.
\end{eqnarray*}
Hence
$$
P_1^{-1} X_\alpha P_1 = i^{\wgt(\alpha)} X_\alpha Z_\alpha.
$$
Finally, for any $\gamma\in\F_q$, $\gamma\ne 0$, we define $P_\gamma$
as
\begin{equation}\label{def:P_even}
P_\gamma := M_{\gamma_0}^{-1} P_1 M_{\gamma_0}
\qquad\text{where $\gamma_0^2=\gamma$.}
\end{equation}
(Note that $\gamma_0$ is uniquely defined as $x\mapsto x^2$ is an
automorphism of $\F_q=\F_{2^m}$.) The matrix $M_\gamma$ is a
permutation matrix, hence $P_\gamma$ is diagonal and commutes with
$Z_\beta$. The action of $P_\gamma$ on $X_\alpha$ is given by
\begin{eqnarray*}
P_\gamma^{-1} X_\alpha P_\gamma
&=&M_{\gamma_0}^{-1} P_1^{-1} M_{\gamma_0} 
   X_\alpha 
   M_{\gamma_0}^{-1} P_1 M_{\gamma_0}\\
&=&M_{\gamma_0}^{-1} P_1^{-1} X_{\alpha\gamma_0} P_1 M_{\gamma_0}\\
&=&M_{\gamma_0}^{-1} i^{\wgt(\alpha\gamma_0)} X_{\alpha\gamma_0} Z_{\alpha\gamma_0} M_{\gamma_0}\\
&=& i^{\wgt(\alpha\gamma_0)} X_{\alpha} Z_{\alpha\gamma_0^2}\\
&=& i^{\wgt(\alpha\gamma_0)} X_{\alpha} Z_{\alpha\gamma}.
\end{eqnarray*}

Summarizing, the operators $P_\gamma$ (defined by (\ref{def:P_odd})
for $q$ odd and (\ref{def:P_even}) for $q$ even) acts on
$(\alpha,\beta)$ always as
$\overline{P}_\gamma:=\left(\begin{smallmatrix}1&\gamma\\0&1\end{smallmatrix}\right)$.

Altogether, we have shown
\begin{theorem}\label{satz:SL2}
The group $J_{\F_q}:=\langle \DFT, P_\gamma, M_\gamma\colon
\gamma\in\F_q^*\rangle$ acts on $\overline{{\cal G}}_1$, the error
group ${\cal G}_1$ modulo the center, as $SL(2,\F_q)$.
\end{theorem}
\proof{%
The action of the matrices $\DFT$, $P_\gamma$, $M_\gamma$ on the error
group $\overline{{\cal G}}_1$ is given by
$$
\overline{\DFT}:=\left(\begin{matrix}0&-1\\1&0\end{matrix}\right),
\quad
\overline{P}_\gamma:=\left(\begin{matrix}1&\gamma\\0&1\end{matrix}\right),
\quad\text{and}\quad
\overline{M}_\gamma:=\left(\begin{matrix}\gamma^{-1}&0\\0&\gamma\end{matrix}\right),
$$
respectively. These matrices generate $SL(2,\F_q)$ (see, e.\,g.,
\cite{OMea78}).
}
As $SL(2,\F_q)$ acts transitively on the non-zero vectors (see,
e.\,g., \cite{Hup67}), we obtain
\begin{corollary}\label{cor:trans}
The group $J_{\F_q}$ acts transitively on the non-trivial elements of
$\overline{{\cal G}}_1$.
\end{corollary}

So far, we have only considered the action of single qudit operations
on the error group ${\cal G}_n$. Our goal is to transform an arbitrary
Abelian stabilizer group ${\cal S}\subseteq {\cal G}_n$ into ${\cal
S}_0$ which corresponds to the code ${\cal C}_0=[\![n,k,1]\!]$.  As
${\cal C}_0$ cannot correct errors and single qudit operations do not
change the error-correcting properties of a stabilizer code, we need
additional transformations which are $\ADD$-gates acting on pairs of
qudits. Combining the definitions of $X_\alpha$ and $\ADD^{(1,2)}$
(see Definition~\ref{def:elemGates}~\ref{def:X} and \ref{def:ADD}), we can
rewrite the $\ADD$-gate as
\begin{eqnarray}
\ADD^{(1,2)}&:=&\sum_{x,y\in\F_q}\ket{x}_1\ket{x+y}_2\bra{y}_2\bra{x}_1
\nonumber\\
&=&\sum_{\alpha\in\F_q}\ket{\alpha}\bra{\alpha}\otimes X_\alpha.
\label{eq:rewriteADD}
\end{eqnarray}
Using (\ref{eq:rewriteADD}), it is easy to show that $\ADD^{(1,2)}$
commutes with all matrices of the form $Z_{\beta_1}\otimes
X_{\alpha_2}$. 

\newpage
For elements of the form $X_{\alpha_1}\otimes Z_{\beta_2}$ we get
\begin{eqnarray*}
\noalign{\noindent$(\ADD^{(1,2)})^{-1}(X_{\alpha_1}\otimes Z_{\beta_2})\ADD^{(1,2)}$}\\
&=&
(\ADD^{(1,2)})^{-1}\sum_{v,w\in\F_q}\omega^{\trace(\beta_2 w)}
  \ket{v+\alpha_1}_1\ket{w}_2\bra{w}_2\bra{v}_1
\sum_{x,y\in\F_q}\ket{x}_1\ket{x+y}_2\bra{y}_2\bra{x}_1\\
&=&
(\ADD^{(1,2)})^{-1}
  \sum_{x,y\in\F_q}\omega^{\trace(\beta_2(x+y))}
  \ket{x+\alpha_1}_1\ket{x+y}_2\bra{y}_2\bra{x}_1\\
&=&
\sum_{v,w\in\F_q}\ket{v}_1\ket{w}_2\bra{v+w}_2\bra{v}_1
  \sum_{x,y\in\F_q}\omega^{\trace(\beta_2(x+y))}
  \ket{x+\alpha_1}_1\ket{x+y}_2\bra{y}_2\bra{x}_1\\
&=&
  \sum_{x,y\in\F_q}\omega^{\trace(\beta_2 x)}\omega^{\trace(\beta_2 y)}
    \ket{x+\alpha_1}_1\ket{y-\alpha_1}_2\bra{y}_2\bra{x}_1\\
&=& (X_{\alpha_1} Z_{\beta_2})\otimes(X_{-\alpha_1} Z_{\beta_2}).
\end{eqnarray*}
This proofs
\begin{lemma}\label{lemma:ADD}
The transformation $\ADD^{(1,2)}$ acts on
$((\alpha_1,\beta_1),(\alpha_2,\beta_2))$ as
$$
\overline{\ADD}^{(1,2)}:= \left(
\begin{smallmatrix}
1 & 0 &-1 & 0\\
0 & 1 & 0 & 0\\
0 & 0 & 1 & 0\\
0 & 1 & 0 & 1
\end{smallmatrix}
\right),
$$
i.\,e., $\beta_2$ is added to $\beta_1$ and $\alpha_1$ is subtracted
from $\alpha_2$.
\end{lemma}

\subsection{Encoding Algorithms}
With this preparation, we can formulate our algorithm to efficiently
compute efficient quantum circuits for the encoding of stabilizer
codes. 
\begin{figure}[tbh]
\rule{\textwidth}{0.5pt}
\begin{tabular}{ll}
{\sc Input:} &a stabilizer matrix $(X|Z)\in\F_q^{(n-k)\times 2n}$\\
{\sc Output:}&a decoding transformation
$D:=F\cdot A_{n-k}T_{n-k} \cdot\ldots\cdot A_2 T_2 \cdot A_1 T_1 $
\end{tabular}\\
\rule{\textwidth}{0.5pt}
\begin{listing}
$L\leftarrow \emptyset$\cr
foreach row $i=1$ to $n-k$ do\label{step:for_i}\cr
\> foreach column $j=1$ to $n$ do\label{step:for_j}\cr
  \>\> if $(X_{ij},Z_{ij})\ne(0,0)$ then\cr
  \>\>\> find a transformation $T_{ij}\in J_{\F_q}$ such that
$(X_{ij},Z_{ij})\cdot \overline{T}_{ij}=(1,0)$\label{step:T_ij}\cr
  \>\>\> foreach row $l=i$ to $n-k$ do\label{step:for_l}\cr
  \>\>\>\> $(X_{lj},Z_{lj})\leftarrow(X_{lj},Z_{lj})\cdot\overline{T}_{ij}$\cr
  \>\>\> end for\label{step:end_for_l}\cr
  \>\> else\cr
  \>\>\> $T_{ij}\leftarrow id$\cr
  \>\> end if\cr
  \> end for\label{step:end_for_j}\cr
  \> $T_i \leftarrow T_{i,1}^{(1)}\otimes
      T_{i,2}^{(2)}\otimes\ldots\otimes T_{i,n}^{(n)}$\label{step:T_i}\cr 
  \> find the first column $l\notin L$ where $X_{il}\ne 0$\label{step:pivot}\cr
  \> include $l$ into $L$\label{step:L}\cr
  \> $A_i\leftarrow id$\cr
  \> foreach column $j=1$ to $n$ do\label{step:for_j2}\cr
  \>\> if $l\notin L$ and $X_{ij}=1$ then\cr
  \>\>\> $A_i\leftarrow A_i\cdot\ADD^{(l,j)}$\cr
  \>\>\> foreach row $\mu=i$ to $n-k$ do\label{step:for_mu}\cr
  \>\>\>\> $(X_{\mu j},Z_{\mu j})\leftarrow(X_{\mu j}-X_{lj},Z_{\mu j}+Z_{lj})$\cr
  \>\>\> end for\label{step:end_for_mu}\cr
  \>\> end if\cr
  \> end for\label{step:end_for_j2}\cr
end for\label{step:end_for_i}\cr
$F\leftarrow \prod_{l\in L}\DFT^{(l)}$\cr
\rule{0pt}{11pt}return $(F,A_{n-k},T_{n-k},\ldots, A_2, T_2, A_1, T_1)$
\end{listing}%
\fcaption{Algorithm to compute a decoding circuit for a qudit
stabilizer code.\label{fig:Encoding}}
\end{figure}%

\begin{theorem}
Let ${\cal C}=[\![n,k,d]\!]_q$ be a stabilizer code for a qudit system
of prime power dimension $q=p^m$ with Abelian stabilizer group ${\cal
S}\subseteq {\cal G}_n$ and stabilizer matrix
$(X|Z)\in\F_q^{(n-k)\times 2n}$.

Then the algorithm shown in Figure~\ref{fig:Encoding} (on
page~\pageref{fig:Encoding}) computes a decoding circuit $D$ for
${\cal C}$ with at most $n(n-k)-\binom{n-k+1}{2}$ $\ADD$ gates and
$O(n(n-k))$ single qudit gates of type $\DFT$, $P_\gamma$, or
$M_\gamma$.

The running time of the algorithm itself is $O(n(n-k)^2)$.
\end{theorem}
\proof{%
We show that the algorithm computes a transformation $D$ such that
$D^{-1} {\cal S} D=_{\pi}{\cal S}_0$ (see
eq.~(\ref{eq:conjugation})). Instead of operating on the stabilizer
group ${\cal S}$ itself, we use the stabilizer matrix
representation. So our goal is to find a transformation $\overline{D}$
such that
\begin{equation}\label{eq:only_Z}
(X|Z)\overline{D}=_\pi (0|A0),\qquad\text{where
$A\in\F_q^{(n-k)\times(n-k)}$ has full rank.}
\end{equation}
In order to obtain a more compact graphical representation of the
final quantum circuit, we will first transform the stabilizer matrix
first into the form
\begin{equation}\label{eq:only_X}
(X|Z)\overline{D}'=_\pi (A0|0),\qquad\text{where
$A\in\F_q^{(n-k)\times(n-k)}$ has full rank,}
\end{equation}
and use $(n-k)$ local Fourier transformations to exchange the $X$- and the
$Z$-part of the stabilizer matrix.

We prove the correctness of our algorithm by induction over $i$,
corresponding to the loop
steps~\ref{step:for_i}--\ref{step:end_for_i}.  The induction
hypothesis is that the first $i-1$ rows of the stabilizer matrix are,
again up to a permutation of the columns, in the form $(A_{i-1}0|0)$,
where $A$ is an $(i-1)\times(i-1)$ matrix of full rank.

After the loop steps~\ref{step:for_j}--\ref{step:end_for_j}, the
$Z$-part of the $i$-th row of the stabilizer matrix will be zero. From
Theorem~\ref{satz:SL2} and Corollary~\ref{cor:trans} it follows that
in step~\ref{step:T_ij} we can always find a transformation
${T}_{ij}\in J_{\F_q}$ with the desired property. The loop
steps~\ref{step:for_l}--\ref{step:end_for_l} updates the stabilizer
matrix.  In step~\ref{step:T_i}, we combine the transformations
applied to each qudit to a transformation on all qudits.  The
definition of the stabilizer matrix (see
Definition~\ref{def:stab_mat}) implies that it has full rank. Hence in
step~\ref {step:pivot}, we will always find a column $l$ with a
non-zero entry $X_{il}$. That column will be recorded in
step~\ref{step:L}. Then the loop
steps~\ref{step:for_j2}--\ref{step:end_for_j2} searches for columns
$j\notin L$ where $X_{ij}$ is non-zero. Applying an $\ADD$-gate with
control $l$ and target $j$ will change those positions $X_{ij}$ to
zero.  The loop steps~\ref{step:for_mu}--\ref{step:end_for_mu} updates
the stabilizer matrix accordingly. So after
step~\ref{step:end_for_j2}, only the entries $X_{il}$ for $l\in L$ may
be non-zero, all other entries in the $i$-th row of the stabilizer
matrix are zero. This completes the induction step from row $i-1$ to
row $i$.

After step~\ref{step:end_for_i}, the $Z$-part of the stabilizer matrix
is zero, and only the $n-k$ columns $l\in L$ of the $X$-part are
non-zero. Applying the transformation $F$, i.\,e., local Fourier
transformations at those positions, yields the stabilizer group ${\cal
S}_0$, completing the proof of the correctness of our algorithm.

The running time of the algorithm follows directly from that fact that
at most three for-loops are nested, two of them iterating over the
$n-k$ rows, one over the $n$ columns.

Each of the $n-k$ transformations $A_j$ is a product of at most $n-j$
$\ADD$-gates.  Hence the total number of $\ADD$ gates is at most
$\sum_{j=1}^{n-k}(n-j)=n(n-k)-\binom{n-k+1}{2}$. Each of the $n-k$
transformations $T_j$ is the product of $O(n)$ single qudit
operations.  Together with the $n-k$ local Fourier transformations
$\DFT$ in the transformation $F$, the number of single qudit gates is
$O(n(n-k))$.
}

Note that, in contrast to the situation of qubits, for $q=p^m$, $p>2$,
the $\ADD$-gate over $\F_q$ for $q=p^m$, $p>2$, is not its own
inverse. So a different graphical representation for $\ADD^{-1}$ is
required. For simplicity, we have not introduced such a
representation. This is also the main reason why we do not directly
compute the transformation $D$ of (\ref{eq:only_Z}), but decompose it
as $D=F\cdot D'$ (see eq.~(\ref{eq:only_X})).

It is not difficult to modify our algorithm such we may save the final
$n-k$ local Fourier transformations. In step~\ref{step:T_ij}, we have
to find a transformation $T_{ij}\in J_{\F_q}$ such that
$(X_{ij},Z_{ij})\cdot \overline{T}_{ij}=(0,1)$. Furthermore, we have
to replace the $\ADD$-gates by inverse $\ADD$-gates. The modified
steps of the algorithm are:

\medskip
\noindent
\begin{tabular}{rl}
5' & find a transformation $T_{ij}\in J_{\F_q}$ such that $(X_{ij},Z_{ij})\cdot \overline{T}_{ij}=(0,1)$\\[1ex]
19'& $A_i\leftarrow A_i\cdot(\ADD^{-1})^{(l,j)}$\\
20'& foreach row $\mu=i$ to $n-k$ do\\
21'& \quad $(X_{\mu j},Z_{\mu j})\leftarrow(X_{\mu j}+X_{lj},Z_{\mu j}-Z_{lj})$\\
22'& end for
\end{tabular}

\section{Example}\label{sec:Example}

We illustrate the algorithm of Figure~\ref{fig:Encoding} using a
quantum code over qutrits. A stabilizer matrix of the code ${\cal
C}=[\![9,5,3]\!]_3$ is given by
$$
(X|Z)=
\left(
\begin{array}{*{9}{c}|*{9}{c}}
1&0&0&2&1&2&2&0&1 & 0&0&2&1&2&2&0&1&1\\
0&1&1&2&0&2&2&1&0 & 0&0&1&2&1&1&0&2&2\\
0&0&2&1&2&2&0&1&1 & 1&0&2&0&0&1&2&1&2\\
0&0&1&2&1&1&0&2&2 & 0&1&2&1&1&0&2&0&2
\end{array}
\right).
$$
In the first step, we transform each pair $(\alpha_i,\beta_i)$ of the
first row that is non-zero to $(1,0)$. This is achieved by the
transformation 
$$
T_1=
id\otimes 
id \otimes 
M_2\DFT \otimes 
P_1 M_2 \otimes 
P_1  \otimes 
P_2 M_2 \otimes 
M_2 \otimes 
\DFT \otimes 
P_2.
$$
The resulting stabilizer matrix is
$$
(X|Z)=
\left(
\begin{array}{*{9}{c}|*{9}{c}}
1&0&1&1&1&1&1&1&1 & 0&0&0&0&0&0&0&0&0\\
0&1&2&1&0&1&1&2&0 & 0&0&1&2&1&1&0&2&2\\
0&0&1&2&2&1&0&1&1 & 1&0&2&2&2&1&1&2&1\\
0&0&1&1&1&2&0&0&2 & 0&1&1&0&2&1&1&1&0
\end{array}
\right).
$$
The first non-zero column is the first one. So using the
transformation
$$
A_1:=\ADD^{(1,3)}\ADD^{(1,4)}\ADD^{(1,5)}\ADD^{(1,6)}\ADD^{(1,7)}\ADD^{(1,8)}\ADD^{(1,9)}
$$
we obtain
$$
(X|Z)=
\left(
\begin{array}{*{9}{c}|*{9}{c}}
1&0&0&0&0&0&0&0&0 & 0&0&0&0&0&0&0&0&0\\
0&1&2&1&0&1&1&2&0 & 0&0&1&2&1&1&0&2&2\\
0&0&1&2&2&1&0&1&1 & 0&0&2&2&2&1&1&2&1\\
0&0&1&1&1&2&0&0&2 & 0&1&1&0&2&1&1&1&0
\end{array}
\right).
$$
Note that the first column of the $Z$-matrix is zero. This follows
from the fact that the corresponding stabilizer elements commute with
the $X^{(1)}$ which corresponds to the first row.

In the next step, we use the transformations
$$
T_2:=
id\otimes id \otimes P_1 M_2\otimes P_1\otimes \DFT\otimes P_2\otimes id\otimes P_2 M_2\otimes M_2 \DFT
$$
and
$$
A_2:=\ADD^{(2,3)}\ADD^{(2,4)}\ADD^{(2,5)}\ADD^{(2,6)}\ADD^{(2,7)}\ADD^{(2,8)}\ADD^{(2,9)}
$$
to obtain
$$
(X|Z)=
\left(
\begin{array}{*{9}{c}|*{9}{c}}
1&0&0&0&0&0&0&0&0 & 0&0&0&0&0&0&0&0&0\\
0&1&0&0&0&0&0&0&0 & 0&0&0&0&0&0&0&0&0\\
0&0&2&2&2&1&0&2&2 & 0&0&0&1&1&0&1&2&1\\
0&0&2&1&2&2&0&0&0 & 0&0&1&1&2&2&1&2&2
\end{array}
\right).
$$
Using the transformations
$$
T_3:=
id\otimes id\otimes M_2\otimes P_1 M_2\otimes P_1 M_2\otimes id\otimes \DFT\otimes P_2 M_2\otimes P_1 M_2
$$
and
$$
A_3:=\ADD^{(3,4)}\ADD^{(3,5)}\ADD^{(3,6)}\ADD^{(3,7)}\ADD^{(3,8)}\ADD^{(3,9)}
$$
yields
$$
(X|Z)=
\left(
\begin{array}{*{9}{c}|*{9}{c}}
1&0&0&0&0&0&0&0&0 & 0&0&0&0&0&0&0&0&0\\
0&1&0&0&0&0&0&0&0 & 0&0&0&0&0&0&0&0&0\\
0&0&1&0&0&0&0&0&0 & 0&0&0&0&0&0&0&0&0\\
0&0&1&1&0&1&0&2&2 & 0&0&0&1&2&2&0&1&1
\end{array}
\right).
$$
For the last row, we use the transformations
$$
T_4:=id\otimes id\otimes id\otimes P_2\otimes M_2\DFT\otimes P_1\otimes id\otimes P_1 M_2\otimes P_1 M_2
$$
and
$$
A_4:=\ADD^{(4,5)}\ADD^{(4,6)}\ADD^{(4,8)}\ADD^{(4,9)}
$$ 
and get
$$
(X|Z)=
\left(
\begin{array}{*{9}{c}|*{9}{c}}
1&0&0&0&0&0&0&0&0 & 0&0&0&0&0&0&0&0&0\\
0&1&0&0&0&0&0&0&0 & 0&0&0&0&0&0&0&0&0\\
0&0&1&0&0&0&0&0&0 & 0&0&0&0&0&0&0&0&0\\
0&0&1&1&0&0&0&0&0 & 0&0&0&0&0&0&0&0&0
\end{array}
\right).
$$
Note that after the first three steps, we have $L=\{1,2,3\}$. Hence
not the third, but the fourth qudit is the control qudit for the
$\ADD$ gates of the last row. Equivalently, we could have subtracted
the third row from the fourth to obtain $X_{4,3}=0$. The stabilizer
group would not have changed as the addition of one row to another
corresponds to multiplying one generator by another.

Finally, a Fourier transformation on the first four positions
transforms the stabilizer group into the form $S_0=\langle
Z_1^{(1)},Z_1^{(2)},Z_1^{(3)},Z_1^{(4)}\rangle$. 

\begin{figure}[hbt]
\centerline{\scriptsize\unitlength0.7\unitlength
\def\DFT{F}
\quad\inputwires[,,,,$\ket{\phi_{\text{enc}}}\left\{\rule{0pt}{90\unitlength}\right.$\kern-2mm](9)
\rlap{\OneQubitGate[6](3,3){$M_2$}\OneQubitGate[6](3,3){$\DFT$}}%
\rlap{\OneQubitGate[5](1,1){$P_1$}\OneQubitGate[4](1,2){$M_2$}}%
\rlap{\OneQubitGate[4](1,1){$P_1$}}%
\rlap{\OneQubitGate[3](1,1){$P_2$}}%
\rlap{\OneQubitGate[2](1,1){$M_2$}}%
\rlap{\OneQubitGate[1](1,1){$\DFT$}}%
\OneQubitGate(1,1){$P_2$}\OneQubitGate(1,4){$M_2$}%
\rlap{\cnot(1,3,9)}%
\rlap{\cnot(1,4,9)}%
\rlap{\cnot(1,5,9)}%
\rlap{\cnot(1,6,9)}%
\rlap{\cnot(1,7,9)}%
\rlap{\cnot(1,8,9)}%
\cnot(1,9,9)%
\rlap{\OneQubitGate[6](3,3){$P_1$}\OneQubitGate[6](3,3){$M_2$}}%
\rlap{\OneQubitGate[5](1,1){$P_1$}}%
\rlap{\OneQubitGate[4](1,1){$\DFT$}}%
\rlap{\OneQubitGate[3](1,1){$P_2$}}%
\rlap{\OneQubitGate[1](2,2){$P_2$}\OneQubitGate[1](5,5){$M_2$}}%
\OneQubitGate(1,1){$M_2$}%
\OneQubitGate(1,1){$\DFT$}
\rlap{\cnot(2,3,9)}%
\rlap{\cnot(2,4,9)}%
\rlap{\cnot(2,5,9)}%
\rlap{\cnot(2,6,9)}%
\rlap{\cnot(2,7,9)}%
\rlap{\cnot(2,8,9)}%
\cnot(2,9,9)%
\rlap{\OneQubitGate[6](3,3){$M_2$}}%
\rlap{\OneQubitGate[5](1,1){$P_1$}\OneQubitGate[5](4,4){$M_2$}}%
\rlap{\OneQubitGate[4](1,1){$P_1$}\OneQubitGate[4](1,1){$M_2$}}%
\rlap{\OneQubitGate[2](2,2){$\DFT$}}%
\rlap{\OneQubitGate[1](1,1){$P_2$}\OneQubitGate[1](3,3){$M_2$}}%
\OneQubitGate(1,1){$P_1$}\OneQubitGate(1,1){$M_2$}
\rlap{\cnot(3,4,9)}%
\rlap{\cnot(3,5,9)}%
\rlap{\cnot(3,6,9)}%
\rlap{\cnot(3,7,9)}%
\rlap{\cnot(3,8,9)}%
\cnot(3,9,9)%
\rlap{\OneQubitGate[5](4,4){$P_2$}}%
\rlap{\OneQubitGate[4](1,1){$M_2$}\OneQubitGate[4](5,5){$\DFT$}}%
\rlap{\OneQubitGate[3](1,1){$P_1$}}%
\rlap{\OneQubitGate[1](2,2){$P_1$}\OneQubitGate[1](3,3){$M_2$}}%
\OneQubitGate(1,1){$P_1$}\OneQubitGate(1,1){$M_2$}
\rlap{\cnot(4,5,9)}%
\rlap{\cnot(4,6,9)}%
\rlap{\cnot(4,8,9)}%
\cnot(4,9,9)%
\rlap{\OneQubitGate[8](1,1){$\DFT$}}%
\rlap{\OneQubitGate[7](1,1){$\DFT$}}%
\rlap{\OneQubitGate[6](1,1){$\DFT$}}%
\OneQubitGate(1,6){$\DFT$}%
\outputwires[$\ket{0}$,$\ket{0}$,$\ket{0}$,$\ket{0}$,,,$\kern-2mm\left.\rule{0pt}{50\unitlength}\right\}\ket{\phi_{\text{in}}}$](9)
}
\fcaption{Decoding circuit for the ternary quantum code ${\cal
C}=[\![9,5,3]\!]_3$. The Fourier transformation $\DFT$ is abbreviated
as $F$.\label{fig:dec_circuit}}
\end{figure}

The complete decoding circuit is shown in
Figure~\ref{fig:dec_circuit}. Reversing the order of the gates and
replacing each gate by its inverse yields a quantum circuit for
encoding. 

\section{Conclusions}
The quantum circuits for both CSS codes and general stabilizer codes
over qudit systems of prime power dimension resulting from our
algorithms have a similar structure. They consist of an alternating
sequence of single qudit gates and $\ADD$-gates with the same control
qudit. The total number of gates is always at most quadratic
in the number of qudits. For CSS codes, the single qudit gates are
either Fourier transformations or multiplication gates, which are
trivial for the case of qubits. Furthermore, some optimizations are
possible by interchanging the r\^ole of the two classical codes
involved.

The algorithm of \cite{ClGo97}, respectively the modified version of
\cite{Gra02,Gras02}, to compute quantum circuits for the encoding of
qubit stabilizer codes can also be generalized to non-qubit stabilizer
codes. It turns out that the resulting quantum circuits have again the
same structure and complexity as those presented in this paper. The
approach presented here, however, has the advantage that the
transformation $D$ used to conjugate the stabilizer group ${\cal S}$
to ${\cal S}_0$ can also be used to compute ``encoded gates'' which
preserve the code ${\cal C}$. Conjugating gates from the error-group
which preserve the code ${\cal C}_0$ by $D$ results in transformations
that are also in the error-group and preserve the code ${\cal C}$. As
those transformations consist only of single qudit operations, they
can be used for fault-tolerant quantum computation.

\nonumsection{Acknowledgements} Markus Grassl and Martin R\"otteler
acknowledge the hospitality of the Mathematical Sciences Research
Institute, Berkeley, USA. Part of this work was supported by {\em
Deutsche Forschungsgemeinschaft (DFG), Schwerpunktprogramm
Quanten-In\-for\-ma\-tionsverarbeitung (SPP 1078), Projekt AQUA (Be~887/13)\/}
and the IST/FET project Q-ACTA (IST-1999-10596) funded by the European
Commission.

\nonumsection{References}



\end{document}